\documentclass[aps,prl,longbibliography,amsmath,amssymb,groupedaddress,superscriptaddress,floatfix,twocolumn,showkeys]{revtex4-1}
    \usepackage{graphicx}
    \usepackage{dcolumn}
    \usepackage{bm}
    \usepackage{color}
    \usepackage{xcolor}
    \usepackage{multirow}
    \usepackage{siunitx}
    \usepackage{ulem}
    \DeclareSIUnit{\sqrthz}{\ensuremath{\sqrt{\text{\hertz}}}}
    \usepackage{soul}
    \usepackage{makecell}
    \usepackage{svg}
    \usepackage{latexsym}
    \usepackage{amssymb}

    \newcommand{\minus}{\scalebox{0.65}[1.0]{$-$}}
    
\begin{document}
    
    \title{Single atom trapping in a metasurface lens optical tweezer}
    
    \author{T.-W.~Hsu}
        \address{JILA, National Institute of Standards and Technology and University of Colorado, and
    Department of Physics, University of Colorado, Boulder, Colorado 80309, USA}
    \author{W. Zhu}
       \address{National Institute of Standards and Technology, Gaithersburg, Maryland 20899, USA}
    \author{T. Thiele}
            \address{JILA, National Institute of Standards and Technology and University of Colorado, and
    Department of Physics, University of Colorado, Boulder, Colorado 80309, USA}
    \author{M. O. Brown}
            \address{JILA, National Institute of Standards and Technology and University of Colorado, and
    Department of Physics, University of Colorado, Boulder, Colorado 80309, USA}
    \author{S. B. Papp}
           \address{National Institute of Standards and Technology, Boulder, Colorado 80305, USA}
    \author{A. Agrawal}
           \address{National Institute of Standards and Technology, Gaithersburg, Maryland 20899, USA}
    \author{C. A. Regal}
            \address{JILA, National Institute of Standards and Technology and University of Colorado, and
    Department of Physics, University of Colorado, Boulder, Colorado 80309, USA}

    \date{\today}%
    
    \begin{abstract}
Optical metasurfaces of subwavelength pillars have provided new capabilities for the versatile definition of the amplitude, phase, and polarization of light. In this work, we demonstrate that an efficient dielectric metasurface lens can be used to trap and image single neutral atoms with a long working distance from the lens of 3 mm. We characterize the high-numerical-aperture optical tweezers using the trapped atoms and compare with numerical computations of the metasurface-lens performance.  We predict that future metasurfaces for atom trapping will be able to leverage multiple ongoing developments in metasurface design and enable multifunctional control in complex quantum information experiments with neutral-atoms arrays.
\end{abstract}

    \maketitle
    
\section{Introduction}

   Arrays of single trapped neutral atoms are a burgeoning platform for quantum simulation, computing, and metrology~\cite{labuhn2016tunable,weiss2017quantum,norcia2019seconds}. With ground-up control similar to trapped ions, individual atoms can be prepared and entangled~\cite{zhang2010deterministic,wilk2010entanglement,levine2019parallel}, and increasingly hold promise for scalable quantum computing~\cite{bluvstein2022quantum,madjarov2020high,schine2021long}.  However, practical quantum computing requires substantial advances in reducing error rates and scaling qubit number. One upcoming outstanding challenge for neutral atoms arrays is developing scalable and multi-functional optical components that enable site-selection manipulation of hyperfine states and Rydberg excitations, operate in constrained environments, and achieve low scattering and cross talk.  
   
   In ion trap experiments, long-standing efforts in developing integrated optical components have enabled improved parallelism and addressing capabilities~\cite{streed2011imaging,brown2016co,mehta2020integrated}.  Neutral atoms will require a similar trajectory and have many unique requirements.  For example, control of single neutral atoms relies heavily on optical potentials for trapping, either in lattices or arrays of tightly-focused laser beams, termed optical tweezers.  Development of active components, from acousto-optic devices to spatial light modulators, are important for moving and addressing individual atoms~\cite{kim2019large,barredo2018synthetic}.  Static components that reduce reliance on large conventional optics for trapping and focusing will also reduce constraints in increasingly complex vacuum chambers and improve scalability.  Advancing these specialized optical systems will benefit from connection to the forefront of integrated photonics development. 

\begin{figure*}[!t]
        \centering
        \includegraphics[width=\textwidth]{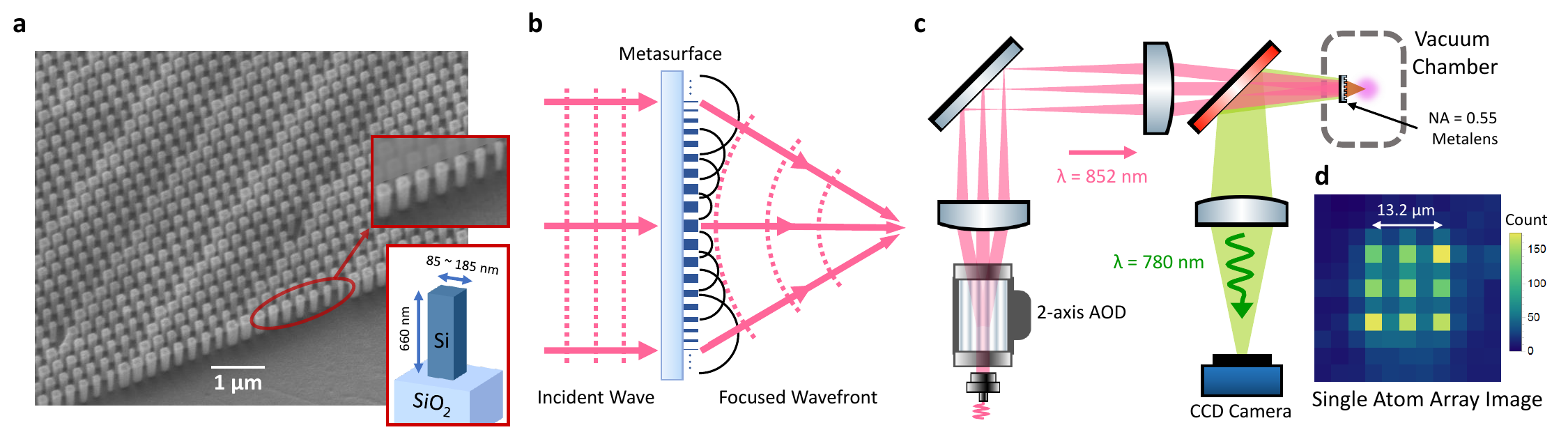}
        \caption{\textbf{Metasurface optics for optical tweezer trapping} (a) Scanning electron micrograph of the fabricated metasurface lens containing a periodic array (lattice constant = 280 nm) of amorphous-Si nanopillars (height 660 nm) of width ranging from 85 nm to 185 nm (dark blue) on top of a 500 $\mu$m thick fused-silica substrate (light blue). Inset shows the varying nanopillar width to achieve the desired phase shift (see Appendix Fig.~\ref{fig:figS1}a). (b) Notional illustration of metasurface lens operation showing light propagation (pink), wavefronts (dashed lines), and secondary wavelets (black semicircles) re-emitted by the nanopillars that interfere to create the focusing wavefront. (c) Optical setup for trapping (pink) and fluorescence imaging (green) of single atoms in an array created with multiple input beams generated using a 2-axis acousto-optic deflectors. (d) Image of a trapped $^{87}$Rb array created by averaging over multiple experiment iterations (100 in this case) with $\sim52\%$ probability of a single atom in each trap per image. The variation in the averaged intensity is caused by trap depth and shape variations that affect relative loading probability and imaging signal in the array.
        }
        \label{fig:fig1}
    \end{figure*}

   Metasurfaces are planar photonic elements composed of a periodic array of subwavelength dielectric or metallic nanostructures that have made significant impact on photonic systems in recent years. Contrary to traditional optical elements that rely on refraction for phase shift, the nanostructures constituting a metasurface couple, resonantly or off-resonantly, and re-radiate the incoming light with a transformed phase, polarization, and amplitude determined by the nanostructure shape, size, and material composition~\cite{kamali2018review}. Electromagnetic modeling, device optimization and fabrication of nanostructures with unprecedented complexity and resolution have enabled multi-functional control of the optical wavefront~\cite{chen_flat_2020,kamali2018review}. By spatially varying the constituent nanopillar width in a pattern similar to a Fresnel lens the re-radiated light can converge at the far field to form a focal point, i.~e.~create a metasurface lens (Fig.~\ref{fig:fig1}).
   
   An important performance metric for utilizing metasurface optics is transmission efficiency, which is governed by multiple factors, including the choice of low-loss and high-index dielectric thin films and the ability to pattern high-aspect-ratio nanostructures with high-fidelity.  Recently, metasurface lenses have been reported with efficiencies ranging from 60$\%$ to 92$\%$~\cite{khorasaninejad_metalenses_2016, chen_gan_2017, zhang_low-loss_2020}, utilizing a range of materials such as TiO$_2$, HfO$_2$, Si and GaN, and operating from the deep ultraviolet to the near infrared.  Further, theuse of inverse design, machine learning, and  multilayers can improve the performance and expand the versatility of metasurface optics~\cite{elsawy_numerical_2020,lin_computational_2021}. 
   
It is informative to compare the spatial wavefront control of metasurfaces to, for example, commercial spatial light modulator (SLM) technologies based on liquid-crystals (LCs) or digital micromirror devices (DMDs). LC-SLMs and DMDs have been used in combination with conventional high-numerical-aperture (high-NA) lenses in atom-array experiments to create arbitrary reconfigurable potentials through spatial modification of the optical wavefront using pixels larger than the optical wavelength. Metasurfaces in contrast consist of deep-subwavelength pillars and operate based upon a waveguide effect that provides large bend angles that can be used for high-NA optics and aggressive wavefront shaping.  To tailor the wavefront, the pillars have a controlled size, spacing and shape, which further enables capabilities such as polarization~\cite{liu_multifunctional_2021} and wavelength multiplexing~\cite{arbabi_two-photon_2018}. While the metasurface used for atom trapping in this work is a static metasurface, active wavefront shaping using metasurfaces is an area of active research~\cite{yang_active_2022}, and has the potential to yield a novel class of SLMs offering capabilities complementary to counterparts based on LCs or DMDs.

   
In atomic physics, metasurface optics are intriguing to explore given their combination of high performance, multifunctional response, and low form-factor.  Initial explorations in metasurfaces for atomic experiments have been used to create beamshaping and splitting elements in magneto-optical traps~\cite{zhu2020dielectric,mcgehee2021magneto}.  In this work, we open up the use of metasurfaces to optical dipole traps, in the form of tightly-focused optical tweezers, and hence to impact increasingly complex quantum information experiments with neutral atoms.  We use a high-NA dielectric metasurface lens to trap and image single atoms (Fig.~\ref{fig:fig1}) and obtain tight trap confinement.  We form an atom array by combining the metasurface lens with tunable acousto-optic deflectors, and characterize the tweezer foci using the trapped atoms. Our successful trapping is an indication that potential deleterious effects of metasurface optics, for example, scattered light, the presence of undeflected zero-order light, or deformations due to absorption and heating of the lens makes negligible contributions to the trapping performance of large-spacing tweezers. We predict that future optimized photonic metasurfaces that leverage ongoing advances in element design libraries and multi-layer design  will enable advanced future high-NA designs with multifunctional performance.
    
   \section{Requirements of high-NA optical tweezers}
   
   In optical tweezers, high-NA optics are key for the creating trapping potentials, the optical addressing of individual atoms in quantum gate protocols, and imaging the fluorescence of single atoms~\cite{schlosser2001sub-poissonian,bakr2009quantum}. Often multielement objective lenses are required to achieve the requisite performance~\cite{schlosser2001sub-poissonian,bakr2009quantum,kaufman2012cooling}, although single aspheric lenses have also been instrumental in state-of-the-art experiments studying interacting Rydberg atoms~\cite{sortais2007diffraction-limited}.  
   
   Optical tweezer experiments require both low aberrations to achieve tight confinement and a high focusing efficiency to achieve sufficient trap depth for a given trapping power and to efficiently image single atoms. Achromatic properties are needed for simultaneously collecting atom fluorescence, conservative trapping in a far off-resonance trap, and often also the delivery of excitation light that controls the atomic state in individual tweezers~\cite{zhang2010deterministic}.  Broadband operation is especially important for multispecies or molecular optical tweezer experiments~\cite{liu2019molecular,singh2021dual,anderegg2019optical}. Further, arbitrary and clean polarization control is increasingly desired.
   
   A long working distance (WD) is required to allow access for laser cooling beams, to maintain sufficient distance between the lens substrate and atoms in high-energy Rydberg states that are sensitive to surface electric dipoles, and to focus light into a complex vacuum chamber or cryogenic environments~\cite{schymik2021single}. In addition, stability of the optics is crucial, for example, in registration of optical tweezers and lattices or for in-vacuum applications.  Further, perturbations to the trap focus due to multi-beam interference or scattered light need to be minimized, especially if they are not static, as these fluctuations can drive atom motion.
      
\section{Metasurface lens overview}

For the demonstration presented in this work, we use a high-contrast transmission-mode metasurface lens (metalens) with NA of 0.55, a measured focusing efficiency of $58\%$ at design wavelength of 852 nm ($56\%$ for atom imaging wavelength  at 780 nm), and a focal length 3 mm (equivalently a WD of 3 mm for the thin lens) (Fig.~\ref{fig:fig1}).  Using the trapped atoms we measure the Gaussian $1/e^2$ radius (waist) of the focused tweezer spot to be $\textrm{w}_0= $($0.80\pm0.04$) $\mu$m, which is consistent with the NA of the designed lens. Further, we create an array of traps with our focusing metasurface lens by introducing multiple beams with tunable angle derived from an acousto-optic deflector pair, and we demonstrate a field-of-view (FoV) of $\pm$11~$\mu$m ($\pm$0.2$^\circ$) (Fig.~\ref{fig:fig1}d), which is consistent with a full theoretical model of the metalens. The FoV is defined as the distance (angle) at which the size is $10 \%$ larger (Strehl ratio $>$ 0.8).  We are able to observe the atoms by measuring through the same metasurface lens, which is a stringent test of the efficiency of the system. Recently metasurface lens trapping and detection of dielectric nanoparticles has been demonstrated, but note these experiments have not required the efficiencies demonstrated in our work because the dielectric particles have been detected using scattered trap light with a much larger magnitude than atomic fluorescence signals~\cite{shen2021chip}.  

The design wavelength of the lens is the trapping wavelength of $\lambda=852$ nm, which is sufficiently far off resonance for $^{87}$Rb atoms to avoid recoil heating.  The 4 mm $\times$ 4 mm square lens is illuminated with a circular Gaussian beam with a $1/e^2$ radius of 2 mm. The lens is also used to collect fluorescence on the $^{87}$Rb D$2$ line at 780 nm.  Given the singlet properties of the lens and the design space offered by the square-shaped nanopillars used in this work, it is not optimized to be diffraction limited at 780 nm.  The metalens is comprised of a thin-film of amorphous-Si (refractive index, $n=3.62+i0.004$ at $\lambda=852$ nm) deposited and patterned on a fused-silica substrate ($n=1.45$) (Fig.~\ref{fig:fig1}a) (for fabrication details, see the Appendix). The materials used for the metalens are fully ultrahigh-vaccuum (UHV) compatible and can resist UHV baking temperatures without any change of properties. The lens is mounted on a sample holder inside an antireflection(AR)-coated glass cell. 

 \section{Metasurface modeling and characterization}
 
We carried out a full numerical simulation of the expected metalens properties using the finite-difference-time-domain (FDTD) method (see Appendix). The theoretical focusing efficiency, estimated as the fraction of the incident light that is diffracted towards the focal spot, is {78$\%$}.  The loss is expected to be from a combination of reflection ({14$\%$}), light remaining in the 0-th order ({6$\%$}), and light scattered into higher orders ({2$\%$}). 

To optically characterize the fabricated metalens we perform a number of experimental tests of the device used in the atom trapping experiment.  First, to characterize the focal spot, we image a pair of 300-nm-diameter pinholes separated by 6.86 $\mu$m using the metalens.  We find that the lens is diffraction limited at 852 nm (Fig.~\ref{fig:fig2}f) by measuring the imaged point spread function (PSF) and fitting it to a Gaussian to find a waist of 0.72 $\mu$m.  At the focus for 780 nm we find a Gaussian waist of 1.1 $\mu$m. Further, the metalens images the 780 nm atom fluorescence out of focus, and we use the pinholes to also analyze and predict the divergence of the imaging system.  Specifically, we find 780 nm with chromatic focal shift of +300 $\mu$m comparing to 852 nm. We also find, as expected for this in-plane square-pillar design, that there is negligible polarization dependence in the focal spot positions.

Second, we characterize the efficiency relevant to both trapping light throughput at 852 nm and collection efficiency at 780 nm.  We assess the combined loss from all factors by measuring the optical tweezer power transmitted through a 300 $\mu$m diameter spatial filter, and also measure the zeroth-order transmission contribution directly (see Appendix).  The measured focusing efficiency, defined as the ratio of power that passes through the 300 $\mu$m spatial filter placed at the lens focus to the total power incident on the substrate, is determined to be 58$\%$ at 852 nm and 56$\%$ for 780 nm, somewhat smaller than the theoretical value. We find zeroth-order light transmitted through the lens (which is conveniently used for system alignment) to be {13$\%$}, somewhat larger than the theoretical estimation. The reduction of the overall efficiency and increase of zeroth-order light in comparison to theory are likely due to fabrication imperfections resulting in nonideal nanopillar cross sections and sidewall tapering.

The amount of zeroth-order undiffracted light can potentially be an issue if it is large enough to alter the trapping potential by interfering with the focused light.  However, from the efficiency measurement of our tightly focused optical tweezers, the intensity at the focused tweezer spot is more than eight orders of magnitude larger than the zeroth-order intensity at trap center.  Hence, the amplitude ratio, which is important for interference effects, is calculated to be $2\times10^4$ times smaller. In the future, the zeroth-order light contribution can be reduced by approaching the theoretical number through better fabrication, or intentionally diverging or deflecting the zeroth-order light in the design.

 \section{Metasurface lens atom trapping}

Atoms are captured into the optical tweezers by overlapping the focus of the metalens with a magneto-optical trap (MOT) and applying polarization-gradient cooling (PGC) for 15 ms while the optical tweezer traps are on~\cite{hsu2020atom}. Light-assisted collisions are used to ensure that only one atom remains in the trap~\cite{schlosser2001sub-poissonian}. To image the atoms, we use a 1 mm diameter probe beam that avoids scattering off of the metasurface by propagating parallel to the substrate (see Fig.~\ref{fig:figS2}e). This beam, in PGC configuration, illuminates the atoms for 25 ms, the fluorescence is collected by the metalens, and the slightly diverging fluorescence is reflected by a dichroic mirror, passed through an imaging lens system and focused onto a charge-coupled device (CCD) camera (see Fig.~\ref{fig:fig1}c and Appendix Fig.~\ref{fig:figS2}). Figure~\ref{fig:fig1}d shows an example single-atom array averaged over 100 loading iterations.

    \begin{figure}
        \includegraphics[width=84mm]{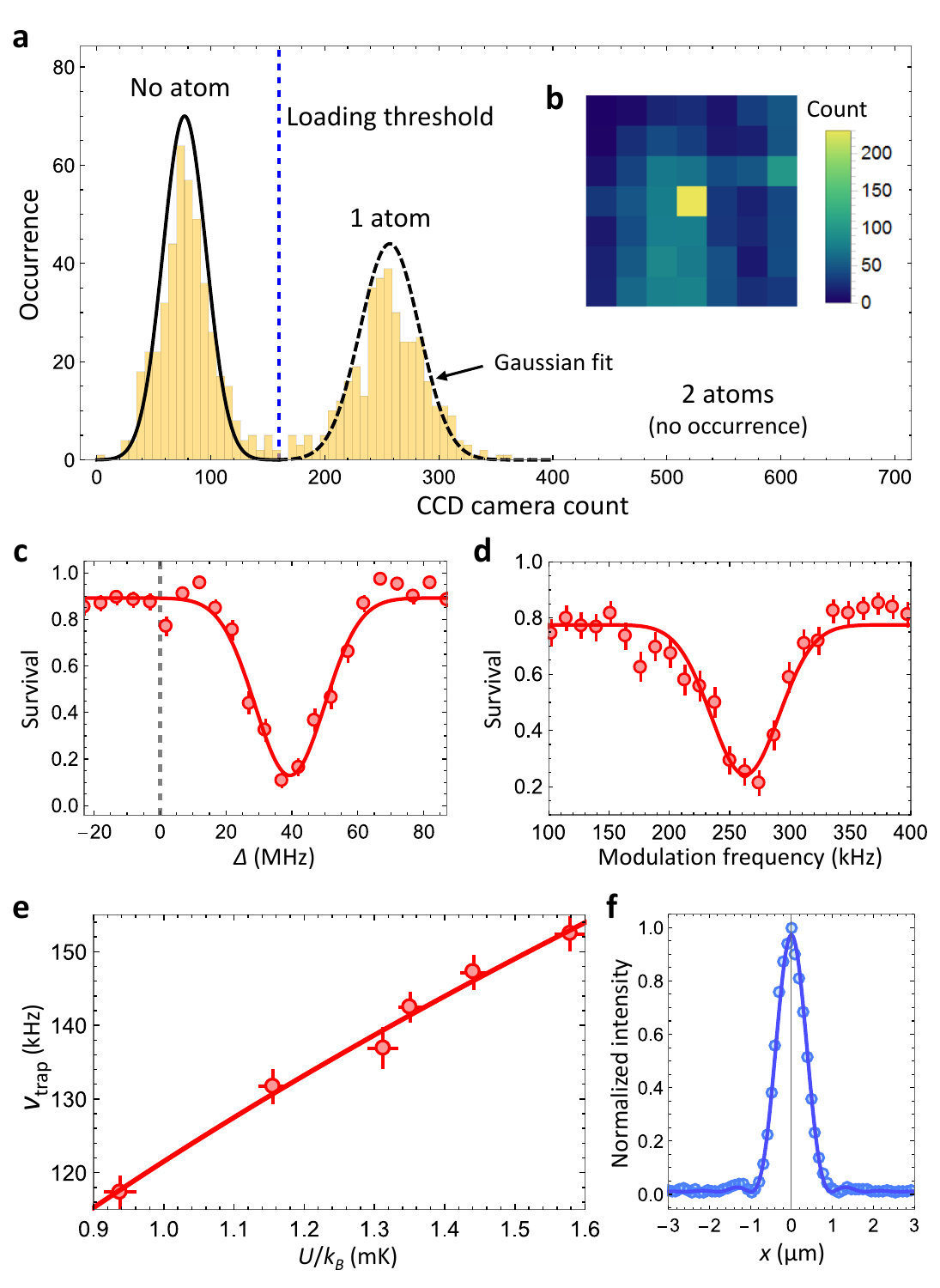}
        \caption{\textbf{Single atom trapping in a metalens optical tweezer} (a) Camera count histogram indicating presence of either 0 or 1 atoms in a tweezer trap. A threshold is chosen to determine if an atom is trapped and to calculate the loading efficiency. (b) Typical single shot fluorescence image of a single atom imaged through the metalens with PGC imaging. (c) Typical light-shift measurement with a Gaussian fit (red line) to the shifted atomic resonance. Dashed line corresponds to free-space $^{87}$Rb $D2$ $F=2$ to $F'=3$ transition (d) Typical parametric heating measurement with a Gaussian fit (red line) to extract the trap frequency ($\nu_{\textrm{trap}}$) from the modulation resonance ($2\nu_{\textrm{trap}}$). Each point is an average of 100 trap loading sequences. (e) Measured trap frequency versus trap depth (light shift)  obtained from multiple measurements similar to (c) and (d). The solid red line is a model fit (see main text) to extract the effective Gaussian tweezer waist seen by the trapped atom. (f) Peak-normalized cross-section of intensity transmission through a 300 nm diameter pinhole imaged at the 852 nm trapping wavelength by the metalens. Solid blue line is an Airy function fitted to the data to extract the spot size and effective NA. (Error bars in (c,d) represent the standard deviation and error bars in (e) are standard error of the fitted Gaussian centers.)
        }
        \label{fig:fig2}
    \end{figure}  
    
We first analyze in detail one trap that is at the center of the metalens FoV. We plot a histogram of the fluorescence counts collected through the metalens and registered on the camera (CCD counts) versus occurrence from a single pixel at the atom location (Figs.~\ref{fig:fig2}a,b). The lower count histogram peak corresponds to background signal with no atom and the second higher CCD count peak corresponds to the fluorescence of the single atom. Collisional blockade prevents loading of more than one atom into the tweezer, as reflected in the absence of a two-atom peak~\cite{schlosser2001sub-poissonian}. We find a loading probability of (47$\pm$5)$\%$. However, due to the limited imaging beam geometry (see Appendix Fig.~\ref{fig:figS2}e), the atom loss during imaging is (10$\pm$2)$\%$. Taking this into account, a loading probability of (52$\pm$5)$\%$  is comparable to typical loading efficiency from other optical tweezer experiments~\cite{lester2015rapid}. We determined the length of time a single atom remains trapped in the optical tweezer focus, with no cooling light present, by holding the atom with variable time between two consecutive images. The measurement gives a lower bound of exponential decay lifetime of 10 sec; atom lifetime assessment in a metalens trap beyond this scale will require additional investigation of background gas collision rates due to finite vacuum level and potential atom loss contributions due to inelastic light scattering from residual trapping light.
    
Next, we characterize the effective tweezer focus size by measuring both the trap depth and the trap frequency (harmonic oscillator strength of atom moving in the optical tweezer light). The measurements are made by determining atom survival following perturbations that depend upon a parameter of interest. For measuring the trap depth $U$, we make use of the fact that the trap light induces an AC-Stark effect that shifts the atomic resonance by 28.8 MHz/mK compared to that in free-space, and we determine the frequency at which resonant light heats the atom out of the trap.   For trap frequency measurements, we modulate the trap between 5$\%$ and 10$\%$ of its depth around the nominal value to parametrically heat at twice the trap frequency ($\nu_{\textrm{trap}}$) and subsequently lower the trap depth to eject hot atoms. Figure~\ref{fig:fig2}c,d show the typical light shift and trap frequency measurements. The trap waist can be deduced from the slope of a graph that plots the trap frequency versus depth as per  $ \nu_{\textrm{trap}}(U,\textrm{w}_0) = \frac{1}{2 \pi}\sqrt{\frac{4 U}{\textrm{w}_0 m_{\textrm{Rb}}}}$ (Fig.~\ref{fig:fig2}e). We extract a $1/e^2$ Gaussian radius of $\textrm{w}_0= $($0.80\pm0.04$) $\mu$m at 852 nm, which is consistent with the value determined from the optical lens characterization (Fig.~\ref{fig:fig2}f). With the clipped Gaussian beam illumination used for the optical tweezer trapping (versus uniform illumination during characterization) we expect the tweezer to have a waist of 0.78 $\mu$m, consistent with the measured value.

\begin{figure}
        \centering
        \includegraphics[width=85mm]{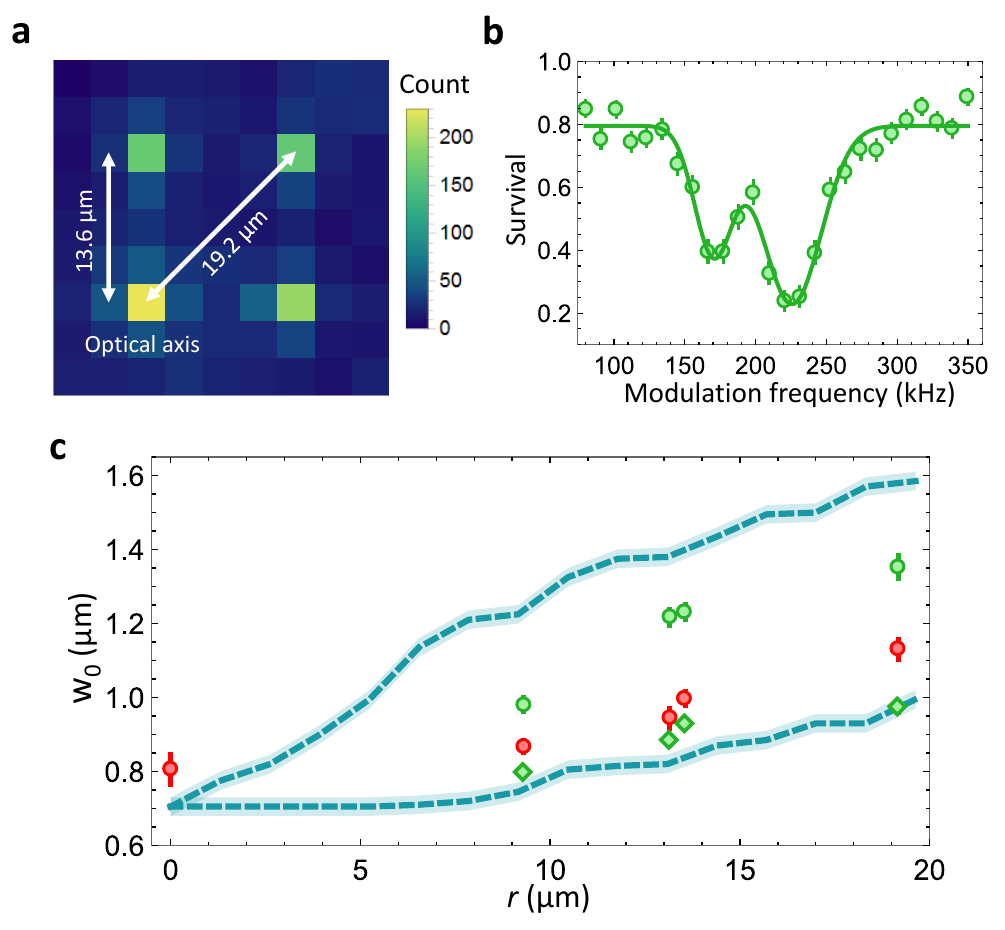}
        \caption{\textbf{Atom array and metalens field of view} (a) PGC fluorescence image of atom array trapped with metasurface optical tweezers. Image is averaged over 100 experimental cycles. Bottom left tweezer is on optical axis of the metalens. The off-axis tweezer site typically have a lower loading probability and non-optimal PGC imaging detuning resulting in a dimmer single atom signal. (b) Example of typical trap frequency measurement data at $\approx$13.6~$\mu$m from FoV center where asymmetric aberrations in the trap are present, along with a double Gaussian (green line) fit. (c) Extracted Gaussian waist as determined from the atom trapping as a function of distance (\textit{r}) to the metalens optical axis (center of FoV). The average waist extracted from a single Gaussian fit to the trap frequency data (red), and the major waist (green circle) and minor waist (green diamond) extracted from data similar to (b) when the two trap frequencies are distinguishable. We compare to theory of the major and minor Gaussian waist fitted from FDTD simulation (see Appendix Fig.~\ref{fig:figS1}c). (Error bars in (b) represent the standard deviation and error bars in (c) are standard error.)
        }
        \label{fig:fig3}
    \end{figure}

An important metric for creating and imaging large atom arrays is the lens FoV. Figure~\ref{fig:fig3} illustrates a study of the metalens tweezer off axis. For this, we create four traps with the lower left tweezer at the center of the FOV (optical axis), and characterize the traps (with various spacing) in analogy to Fig.~\ref{fig:fig2}c,d. In the presence of aberrations the traps become asymmetric, resulting in non-degenerate frequencies in the radial dimensions of the trap. This will manifest as a double-peak structure in the trap frequency measurement (Fig.~\ref{fig:fig3}b). We characterize the FoV by plotting the waist determined from the trap frequency and depth measurements as a function of the distance from the optical axis (Fig.~\ref{fig:fig3}c) and find the aberrations are consistent with FDTD calculations of tweezer intensity from our metalens optical field distribution (blue lines, Fig.~\ref{fig:fig3}c). Here FoV is defined as the distance to a point where the average waist is $10 \%$ larger (Strehl ratio $>$ 0.8) than at the center, and we find a FoV of $\pm11$ $\mu$m ($\pm$0.2$^\circ$).  

     \section{Comparative and future potential}
     
As one comparison, we discuss the performance of a typical commercial asphere that has been used in optical tweezer experiments. Reference~\cite{sortais2007diffraction-limited} uses an aspheric lens with NA = 0.5, a working distance of 5.7 mm, a focal shift of  -40 $\mu$m from 852 nm to 780 nm, and a focal length of 8 mm.  This aspheric lens has a transverse field of view of $\pm 25$  $\mu$m ($\pm$0.18$^\circ$), an inferred beam waist of 1 $\mu$m for the trapping wavelength, and a 0.9 $\mu$m waist for the imaging wavelength. The metasurface studied here has a worse focal shift than a standard asphere, but as discussed below this was not of primary concern in our experiments given the prospects for future control, for example, using wavelength polarization multiplexing.  The singlet metasurface here achieves similar or better performance as the representative asphere for focal length to FoV ratio, i.~e.~angle, and effective NA.

In comparison, a complex high-NA objective lenses used for atom trapping and imaging can have FOV of a few 100 $\mu$m ($\approx\pm$3$^\circ$) combined with achromatic operation over a wide range of wavelengths~\cite{bakr2009quantum,norcia2019seconds}.  While the singlet metalens described in this work does not yet achieve these metrics, we now discuss the horizon of prospects for design features of future metasurfaces. 

As discussed previously, with a metasurface it is possible to achieve a focusing response that is either polarization selective~\cite{hu_all-dielectric_2020} or one that transforms the polarization~\cite{chen_highly_2021}, which are functions not offered by traditional optical lenses.  For example, polarization multiplexing provides a method to trap and collect fluoresence at the diffraction limit for two different wavelengths using a singlet metasurface lens,  and may find utility in combining additional multifunctional beams in complex trapping experiments.  To illustrate this prospect, we have designed and optically tested a sample with in-plane rectangular shape pillars that achieves equal focal lengths for 780 nm and 852 nm light of orthogonal polarization (see Appendix).  This concept can be used to trap at 852 nm, and collect fluoresence at 780 nm, with a 50$\%$ efficiency due to the random polarization of the scattered light from atoms.

More functionality can be achieved with expanding the number of surfaces offered in the design.  To focus on FoV as one metric, an enhanced FoV up to $\pm$25$^\circ$ has been achieved by patterning both sides of the substrate to create a double-layer metasurface~\cite{groever_meta-lens_2017}. We estimate that by using design components similar to the singlet lens presented here, expanding to a doublet can improve the field angle to beyond $\pm$5$^\circ$ at 0.55 NA.

Further design improvements can be achieved through the use of an expanded unit-cell library to include cross, donut and other exotic nanopillar shapes~\cite{shrestha_broadband_2018} or via inverse design~\cite{phan_high-efficiency_2019}. Choosing optimal materials and designs that are robust to nanofabrication variation is expected to offer higher efficiencies that exceed that achieved in the experiments presented here~\cite{chen_flat_2020}. Further, a hybrid lens design consisting of a curved refractive surface and a metasurface patterned on the same substrate will offer additional prospects for enhanced design space~\cite{chen_broadband_2018, nikolov_metaform_2021}. 

   \section{Acknowledgements} This work was supported by the DARPA A-PhI program under Grant No.~FA9453-19-C-0029, ONR Grant No.~N00014-17-1-2245 and N00014-21-1-2594, and  NSF  QLCI  Award OMA - 2016244, and NSF Grant No.~PHYS 1914534. We acknowledge helpful input from Christopher Kiehl and Zhenpu Zhang and technical expertise from Yolanda Duerst.

    \section{Appendix} \label{appendix}
    
    \begin{figure*}[!t]
        \centering
        \includegraphics[width=\textwidth]{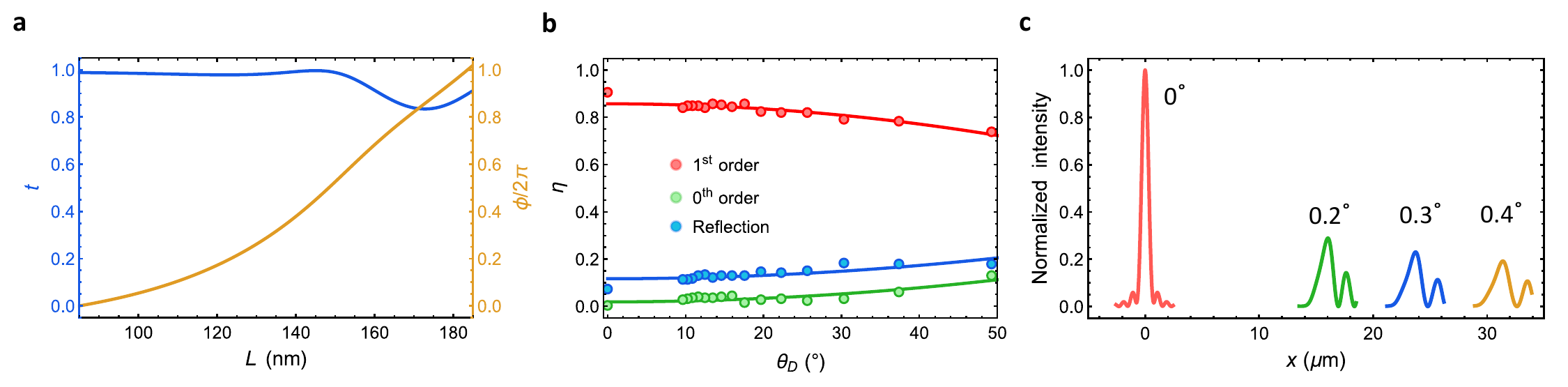}
        \caption{\textbf{Metalens design and simulations} (a) Transmittance $t$ and phase shift $\phi$ as a function of nanopillar side length $L$. (b) Calculated deflection efficiency $\eta_1$, the fraction of light in the undeflected 0th-order $\eta_0$, and the reflectance $\eta_\textrm{refl}$ of aperiodic metasurface beam deflectors as a function of deflection angle $\theta_\textrm{D}$. Circles are data from RCWA simulations and solid lines are parabolic fit. (c) FDTD simulated beam profiles of the focal spots as a function of the angle of incident light.
        }
        \label{fig:figS1}
    \end{figure*} 
    
    \subsection{1.~Metasurface phase profile}        
        
  The metalens used in this study consists of a square lattice (lattice constant $a$ = 280 nm) of a-Si nanopillars on a fused-silica substrate. Each nanopillar unit cell, of in-plane square cross-section (side length $L$) and height $H$ = 660 nm, acts as a phase-delay waveguide (Fig.~\ref{fig:fig1}a). The in-plane side lengths $L$ of the nanopillars vary between 85 nm to 185 nm, forming a library of metasurface unit-cell designs offering transmittance close to unity and relative phase shift $\phi$ covering the full 0 to 2$\pi$ span (Fig.~\ref{fig:figS1}a). This design library is then used to implement the phase profile of an ideal lens~\cite{chen2018broadband}, given by:
   \begin{equation}\label{eq:1}
       \phi(x,y)= \frac{2\pi}{\lambda} \left( f-\sqrt{x^2+y^2+f^2} \right),
   \end{equation}
   where $\lambda$ is the design wavelength (here, $\lambda=852$ nm), $f$ is the focal length (here, $f=3$ mm), and $x,y$ are the coordinates of sampled phase position relative to the lens center.
    
    \subsection{2.~Metasurface theoretical estimation of efficiencies}
    We use the grating averaging technique~\cite{arbabi2020increasing} to estimate the focusing efficiency, the fraction of incident light remaining as undeflected zeroth-order light, and the total reflectance of the mm-scale diameter metalens. Following this technique, we approximate the metalens as a combination of aperiodic beam deflectors. For an aperiodic beam deflector with a deflection angle $\theta_\textrm{D}$ ($\theta_\textrm{D}=$ sin$^{-1} (\lambda/Na)$, where $N$ was chosen to calculate $\theta_\textrm{D}$ between 0$^\circ$ to 50$^\circ$), the deflection efficiency $\eta_1$, the fraction of light in the 0th-order $\eta_0$, and the reflectance $\eta_\textrm{refl}$, for unpolarized input light, are calculated (circles in Fig.~\ref{fig:figS1}b) using rigorous coupled wave analysis (RCWA), and fitted with parabolic functions (solid lines in Fig.~\ref{fig:figS1}b). Finally, the focusing efficiency of the metalens $T_1$, the total fraction in the undeflected 0th-order $T_0$, and the total reflectance $T_\textrm{refl}$, are estimated as the area average of $\eta_1$, $\eta_0$, and $\eta_\textrm{refl}$, respectively, using:
    \begin{equation}\label{eq:2}
        T_i=1/({\pi}R^2)\iint_S \eta_i \,ds
        =2/{R^2}\int_{0}^{R} \eta_i(r)r \,dr,
    \end{equation}
    where $i$= 1, 0, or refl; $r=f$tan$\theta_\textrm{D}$;  and $R$ is the radius of the metalens.

   \subsection{3.~Metasurface theoretical estimation of FoV}
    The beam waist at the focal spot as a function of the distance from the metalens optical axis or, equivalently, the incident angle of the input beam, is calculated using FDTD technique, with a minimum mesh size of 4 nm. Due to the millimeter scale size of the metalens, a cylindrical metalens is simulated instead, implemented by using one unit-cell along the $y$-axis with periodic boundary condition. All the unit cells along the $x$-axis are included in the simulation, and the oblique incidence angle is only applied along the $x$-direction. For a given incident angle, a near-field monitor records the electric and magnetic fields of the output beam at a distance of 50 nm from exit surface of the metasurface. A near-field to far-field projection is then used to calculate the focal spot intensity profile at the focal plane (Fig.~\ref{fig:figS1}c). The major and minor waists of the focal spot are obtained as the distance from the intensity peak to the 1/$e^2$ of peak intensity along the $x$-axis.
    
    \subsection{4.~Metasurface fabrication details}  
    The metasurface optics is fabricated by depositing a layer of 660 nm thick a-Si on a 500 $\mu$m thick fused silica wafer using plasma enhanced chemical vapor deposition (PECVD). A 300 nm thick layer of electron beam resist (ZEP 520A) followed by a 20 nm thick layer of anti-charging conductive polymer (ESpacer 300Z) are spin-coated onto the a-Si film. A 100 keV electron beam lithography system is used to expose the nanopillar pattern, followed by ESpacer removal with deionized water at room temperature, and resist development with hexyl acetate at 4 °C. The developed pattern in the resist layer is transferred to an electron-beam-evaporated 70 nm thick Al$_2$O$_3$ layer using the lift-off technique. By using the patterned Al$_2$O$_3$ layer as an etch mask, inductively-coupled-plasma reactive ion etching (ICP-RIE, gas mixture: SF$_6$ and C$_4$F$_8$; ICP power: 1750 W; radio frequency (RF) power: 15 W) is performed to etch the underlying a-Si layer at 15 °C, to create high-aspect-ratio a-Si nanopillars. The metasurface optics fabrication is finalized by soaking the wafer in a mixture of hydroperoxide and ammonia hydroxide solutions (80 °C for 30 min) to remove the Al$_2$O$_3$ etch mask and any etch residue. 
    
    \subsection{5.~Metasurface optical characterization}
    
    To verify the lens is diffraction limited at 852 nm, we image a pair of pinholes spaced by 6.86 $\mu$m and 300 nm in diameter with the metalens. The pinholes are sufficiently small to be treated as point sources. The magnification of the system is calibrated by using the known separation of the pinholes. Fitting an Airy function to the imaged PSF, a $1/e^2$ Gaussian waist of (0.72$\pm$0.02) $\mu$m and an effective NA of 0.55$\pm$0.01 is extracted (Fig.~\ref{fig:fig2}f), which is consistent with the diffraction limit.
    
    To measure the focusing efficiency, a spatial filter is used to exclude the zeroth order transmission from the focused light. A collimated 852 nm laser beam of 4 mm in diameter illuminates the metalens. A pinhole of dimensions that allow the focused light to be transmitted (300 $\mu$m pinhole of 300 $\mu$m thickness) is then placed at the metalens focus.  A power meter is placed 7 mm away from the metalens (4 mm from the metalens focus), and the pinhole is translated in \textit{x}, \textit{y} and \textit{z} to maximize the power transmitted. The input power and transmitted power are compared to extract the focusing efficiency. The procedure is then repeated for 780 nm and for other input polarizations. The focusing efficiency is found to be 58$\%$ at 852 nm and 56$\%$ for 780 nm and insensitive to polarization rotation for both wavelengths. 
    
\subsection{6.~Sample mounting and vacuum chamber}

\begin{figure*}
        \centering
        \includegraphics[width=\textwidth]{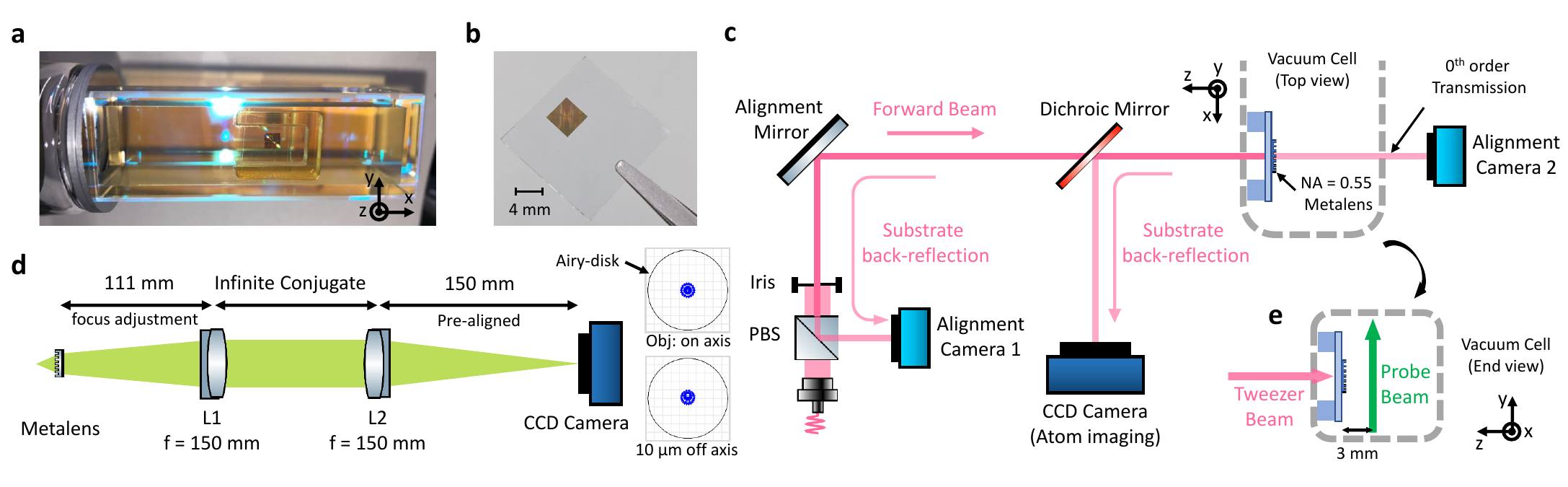}
        \caption{\textbf{Metalens in-vacuum mounting and tweezer alignment} (a) Photo of the metalens sample optically contacted onto the wedged fused silica sample holder that is epoxied onto the AR coated glass cell. (b) Fabricated NA of 0.55 metalens sample designed for 852 nm tweezer light. (c) Schematic illustrating how the tweezer light and science CCD camera are aligned to the metalens sample via substrate back-reflection. (d) Schematic illustration of optical tweezer imaging path with lenses that compensate for out-of-focus imaging due to chromatic focal shift introduced by the metalens. Insets are the ray-tracing simulation of the imaging system (object on-axis and 10 $\mu$m off axis) assuming metalens only has chromatic focal shift and no other aberration. The result shows L1 and L2 does not introduce additional aberrations. Black circle is the diffraction limited Airy-disk. (e) End view of the vacuum cell, showing the probe beam (also resonant heating beam) orientation in relation to metalens sample and tweezer beam. Probe beam is 1 mm in diameter, shines vertically up and is 3 mm away from the metalens overlapping with the optical tweezer focus.
        }
        \label{fig:figS2}
    \end{figure*}

    The metasurface sample is mounted in a Pyrex cell (science cell) with anti-reflection coating on the outside (Fig.~\ref{fig:figS2}a). A sample holder machined from a fused-silica wedge (0.5$^\circ$) with faces polished to better than $\lambda/8$ is epoxied to the inside of the cell with ultra-low outgassing high-temperature epoxy (Epotek-353ND). The epoxy absorbs any minor thermal expansion mismatch between the Pyrex and the fused-silica substrate. The metalens sample (Fig.~\ref{fig:figS2}b) is then optically contacted to the sample holder (Fig.~\ref{fig:figS2}a). The optical contact bonding ensures the metalens substrate remains optically flat after ultra high vacuum (UHV) bake (up to 220 $^\circ$C). The adhesive-free optical contact also allows the cell to be reused indefinitely.  The materials used for the metalens (a-Si and fused-silica) are UHV compatible and can be baked to high temperature ($>$200 $^\circ$C).

    The atomic source is  a magneto-optical trap (MOT) glass cell that is located 40 mm from the science cell and connected through a differential pumping orifice with vacuum conductance of 0.05 L/s. The science cell connects to an ion pump with pumping speed of 25 L/s resulting in a vacuum environment of $<10^{-10}$ hPa measured at the ion pump. A valve between the source MOT cell and the rest of the system isolates the source MOT while the system is vented for sample exchange. The compact construction of the vacuum chamber allows the chamber to be moved elsewhere for sample mounting and UHV baking.
    
    \subsection{7.~Acousto-optic deflectors}
    
    To generate an array of optical tweezers a $1/e^2$ waist = 2 mm collimated beam at 852 nm (pink shaded beams in Fig.~\ref{fig:fig1}c) is launched into a two-axis acousto-optic deflector (AOD) (AA Opto-electronics Model: DTSXY-400-850.930-002). This produces a series of deflected beams with variable angle controlled by the AOD input RF frequencies. This array of angled collimated light is then imaged with a 1:1 relay lens onto the back aperture of the metalens substrate. The relay lens ensures all the deflected beams coincide on the metalens to minimize asymmetric beam clipping.
    
\subsection{8.~Metalens and CCD camera alignment}

    To ensure optimal tweezer performance from the high NA metalens the input light is aligned perpendicular to and centered on to the metalens (Fig.~\ref{fig:figS2}c). The back-reflection of the substrate is used to align the tweezer input light.  The tweezer light ($1/e^2$ waist of 2 mm) is passed through a polarizing beam splitter (PBS) and an iris apertures the beam down to 0.5 mm diameter. Alignment camera 1 (Fig~\ref{fig:figS2}c) is placed on the reflection port of the PBS to monitor the back-reflection from the metalens substrate. This iris allows $<$0.25 mrad angular alignment resolution between the input beam and substrate. Alignment camera 2 (Fig.~\ref{fig:figS2}c) is placed behind the glass cell to monitor the zeroth order metalens transmission. The shadow cast by the structure on the metalens allows the input beam to be centered on the metalens. The input beam is walked while monitoring the image on both alignment cameras until the input is both perpendicular and centered. The residual reflection of the back-reflected tweezer beam from the dichroic mirror (Fig.~\ref{fig:figS2}c light pink) is used to align the position of the science camera and the imaging system. Finally a bandpass filter centered at 780 nm (Semrock LL01-780-12.5) is placed in the imaging path to block any residual tweezer light.
        
\subsection{9.~Imaging path compensation}
    Because the metalens is only designed to be diffraction limited at 852 nm, it is important to characterize the imaging performance of the lens at the atomic fluorescence wavelength of 780 nm. To measure the chromatic focal shift, the metalens is illuminated with a collimated tunable laser source and the focused spot is imaged with an apochromatic microscope objective with NA of 0.9. By changing the microscope focus we determine the chromatic focal shift to be +300 $\mu$m between 852 nm to 780 nm.  We then calculate the signal of an atom trapped at 852 nm focus and emitting 780 nm fluorescence diverges with EFL of \minus39 mm after passing through the metalens (Fig.~\ref{fig:figS2}d). To compensate, a lens of EFL=150 mm (L1 in Fig.~\ref{fig:figS2}d, Thorlabs AC254-150-B) is placed 111 mm from the metalens. The combined optical system (metalens $+$ L1) becomes infinitely conjugate so the tube lens (L2 in Fig.~\ref{fig:figS2}d, Thorlabs AC254-150-B) is decoupled from the compensated imaging system. L2 is pre-aligned to the camera, and L1 is translated to focus the imaging system by only adjusting one optical element. The inset of Fig.~\ref{fig:figS2}d shows the ray-tracing simulation of the imaging system for both on-axis and 10 $\mu$m off-axis on the tweezer plane verifying that the compensation lens and tube lens does not introduce aberrations. The ray-tracing simulation does not include aberration inherent to the metalens design.
    
    To characterize the compensated imaging system, the same 300 nm diameter double pinhole is imaged again with the pinhole positioned at 852 nm focus of the metalens and illuminated with 780 nm light. The resulting PSF has a waist of (1.1$\pm$0.07) $\mu$m which is not diffraction limited (due to the metalens spherical aberration at 780 nm) but sufficient for single atom detection, and the effective solid angle for light collection is equivalent to 0.55 NA.
    
\subsection{10.~Loading and detection optical parameters}
    The single atom loading starts with the three-dimensional (3D) science MOT. The atoms from the dispenser in the source cell are cooled in the transverse direction with MOT laser red detuned from $^{87}$Rb $D2$ $F=2$ to $F'=3$ transition (free-space atomic resonance) by 14 MHz and transported to the science  cell via a push laser beam. The collimated atom beam has a flux up to $~10^8$ s$^{-1}$. The science MOT loading lasts 500 ms with a typical MOT size of $3\times10^7$ atoms and a density of $~10^{9}$ cm$^{-3}$. After loading, the source MOT lasers are shut off and the magnetic field gradient is turned off and the MOT lasers are changed to 144 MHz red detuned from free-space atomic resonance to perform PGC with $\sigma_+\operatorname{-}\sigma_-$ configuration for 15 ms. During the PGC the optical tweezer is turned on to load atoms into the tweezer. The typical free-space PGC temperature is between 30 $\mu$K to 50 $\mu$K, and the tweezer trap depth is typically at 1.3 mK during loading. During the PGC loading the laser is red detuned from the atomic resonance resulting in light assisted collision that assures only a single atom is trapped~\cite{schlosser2001sub-poissonian}. 
    
    To image a single atom in the tweezer, we utilize PGC imaging. The PGC configuration with less detuning cools the atom while scattering photons. The trapped atom is illuminated with a 500 $\mu$m waist and 150 $\mu$W PGC/probe beam (beam geometry shown in Fig.~\ref{fig:figS2}e, $\approx$10$I_{\textrm{sat}}$, 47 MHz red detuned from free-space atomic resonance) for 25 ms and the fluorescence is collected by the metalens ($I_{\textrm{sat}}$ is the saturation intensity of $^{87}$Rb $D2$ $F=2$ to $F'=3$ transition). After passing through the metalens, the slightly diverging fluorescence is reflected by a dichroic mirror and passed through the compensation and imaging lens (L1 and L2 in Fig.~\ref{fig:figS2}d) and focused onto a Princeton Instruments PIXIS 1024B CCD camera. The imaging loss rate is higher than typical PGC imaging due to the probe beam being perpendicular to the metalens substrate so no axial cooling is present during PGC imaging. While a full analysis of fluorescence collection efficiency requires calibration of the probe light intensity, trap depth, and imaging path efficiency, we can roughly compare the expected and measured CCD counts. The total fluorescence collected in the experiment is around $3\%$, and the expected efficiency is $4.5\%$ when only accounting for the solid angle at NA 0.55 and the efficiency of the metalens, but not other loss in the imaging path. The data presented are in CCD counts and are not converted to photon count. The intensity variation in the averaged atom array image presented in Fig.~\ref{fig:fig1}d and Fig.~\ref{fig:fig3}a stems from varying trap depths and aberrations that affect both loading and imaging. In the array trapping experiment, the optical power of the traps are equalized to within 5$\%$ relative difference at the Fourier plane in between the relay lens, but due to aberrations the actual trap depth deviates from the on-axis trap.

\subsection{11.~Trap depth and frequency measurement parameters}
    To measure the tweezer trap depth a resonant heating beam is used. Between the two consecutive PGC imaging sequences the heating beam intensity is set to 100 $\mu$W ($\approx$6$I_{\textrm{sat}}$) and is pulsed on for 60 $\mu$s. The pulse time and intensity of the resonant heating beam are carefully chosen such that the atom only heats out near the atomic resonance. The atom survival versus heating beam detuning is recorded by measuring the atom survival between the two PGC images (Fig.~\ref{fig:fig2}c).
        
    To measure the trap frequency, the tweezer trap depth is modulated around the nominal value between the consecutive PGC imaging and the atom survival is recorded as a function of modulation frequency (Fig.~\ref{fig:fig2}d). The modulation depth ranges between 5$\%$ to 10$\%$ of the trap depth and modulation time ranges from 30 ms to 50 ms. 
    
\subsection{12.~Polarization multiplexed metalens}
\begin{figure}
        \includegraphics[width=84mm]{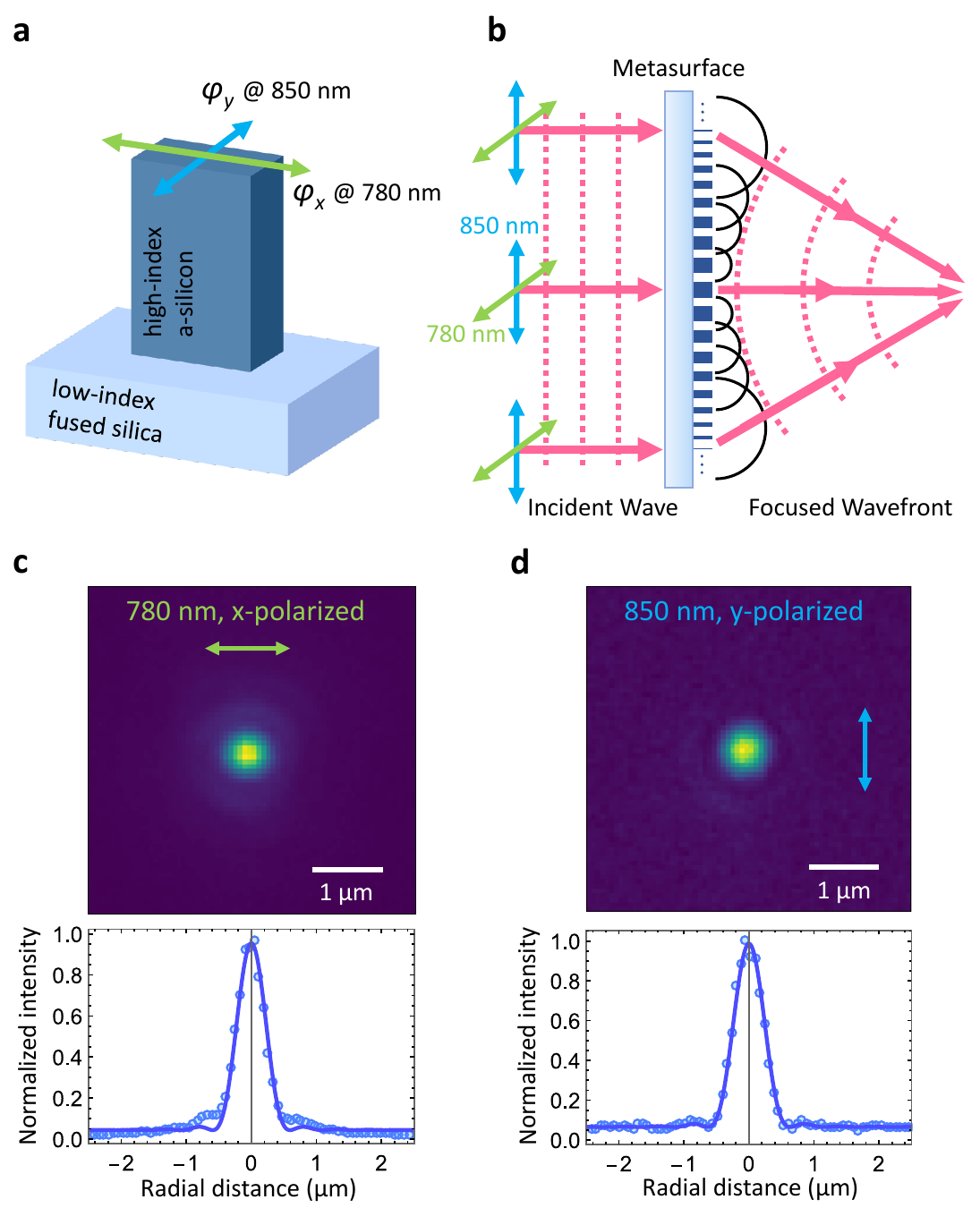}
        \caption{\textbf{Polarization multiplexed metalens} (a) amorphous-Si nanopillar (dark blue) with rectangular cross section on fused silica substrate (light blue) creates phase delay for two orthogonal polarization independently (polarization multiplexing). (b) Illustration of polarization multiplexed metasurface lens operation. Input wavefront (pink dash) with orthogonal polarization for 780 nm (green) and 852 nm (blue) propagates and interact with metasurface. Secondary wavelets (black semicircles) re-emitted by the nanopillars interfere and creates identical focusing wavefront for both 780 nm and 852 nm. (c,d) Experimental PSF focused by the metalens for 780 nm x-polarized and 852 nm y-polarized input light, imaged with 0.95 NA microscope objective without changing the focus. 2D cuts of the PSF shows the fitted Airy function from which the NA is extracted.
        }
        \label{fig:figS3}
    \end{figure}

    To illustrate the flexibility of metalens functionalities, we fabricate and test a polarization-multiplexed metalens that can provide diffraction-limited focusing at both the trapping and fluorescence wavelengths with NA=0.8. This metalens is designed to implement an ideal-lens phase function for $x$-polarized light at $\lambda=780$ nm and for $y$-polarized light at $\lambda=850$ nm, simultaneously, targeting a focal length of 0.5 mm and NA of 0.8, for both wavelengths (Fig.~\ref{fig:figS3}b). This is achieved by assembling the metalens with a library of a-Si birefringent nanopillar unit-cells, each having an in-plane rectangular cross-section that can simultaneously impose the desired local phase shift $\phi_x$ for $x$-polarized light at $\lambda=780$ and $\phi_y$ for $y$-polarized light at $\lambda=850$ nm (Fig.~\ref{fig:figS3}a). The measured intensity distributions at the targeted focal plane reveal diffraction-limited focal spots for both wavelengths (Fig.~\ref{fig:figS3}c and ~\ref{fig:figS3}d). Measured NA for 780 nm $x$-polarized light is 0.8 $\pm$0.01 and 850 nm y-polarized light is 0.82 $\pm$ 0.01. The errors quoted are standard error of fitted Airy function. The focusing efficiencies for $x$-polarized light at $\lambda=780$ and for $y$-polarized light at $\lambda=850$ are 42$\%$ and 45$\%$, respectively. In order to work towards atom trapping in a polarization-multiplexed lens, a lens must be fabricated with more perfect rectangular pillars to slightly improve efficiency and with a larger diameter to increase the working distance. This presents more design and fabrication challenges, and such studies will be a subject of future work.
    
    \end{document}